\documentclass[]{aastex62}

\graphicspath{{./}{figures/}}
\submitjournal{ApJL}

\shorttitle{IceCube Search for Neutrino Emission from Seyfert Galaxies in the Southern Hemisphere}
\shortauthors{IceCube Collaboration}

\begin{document}

\title{\textbf{Evidence for neutrino emission from X-ray Bright Seyfert Galaxies in the Southern Hemisphere using Enhanced Starting Track Events with IceCube}}

\affiliation{III. Physikalisches Institut, RWTH Aachen University, D-52056 Aachen, Germany}
\affiliation{Department of Physics, University of Adelaide, Adelaide, 5005, Australia}
\affiliation{Dept. of Physics and Astronomy, University of Alaska Anchorage, 3211 Providence Dr., Anchorage, AK 99508, USA}
\affiliation{School of Physics and Center for Relativistic Astrophysics, Georgia Institute of Technology, Atlanta, GA 30332, USA}
\affiliation{Dept. of Physics, Southern University, Baton Rouge, LA 70813, USA}
\affiliation{Dept. of Physics, University of California, Berkeley, CA 94720, USA}
\affiliation{Lawrence Berkeley National Laboratory, Berkeley, CA 94720, USA}
\affiliation{Institut f{\"u}r Physik, Humboldt-Universit{\"a}t zu Berlin, D-12489 Berlin, Germany}
\affiliation{Fakult{\"a}t f{\"u}r Physik {\&} Astronomie, Ruhr-Universit{\"a}t Bochum, D-44780 Bochum, Germany}
\affiliation{Universit{\'e} Libre de Bruxelles, Science Faculty CP230, B-1050 Brussels, Belgium}
\affiliation{Vrije Universiteit Brussel (VUB), Dienst ELEM, B-1050 Brussels, Belgium}
\affiliation{Dept. of Physics, Simon Fraser University, Burnaby, BC V5A 1S6, Canada}
\affiliation{Department of Physics and Laboratory for Particle Physics and Cosmology, Harvard University, Cambridge, MA 02138, USA}
\affiliation{Dept. of Physics, Massachusetts Institute of Technology, Cambridge, MA 02139, USA}
\affiliation{Dept. of Physics and The International Center for Hadron Astrophysics, Chiba University, Chiba 263-8522, Japan}
\affiliation{Department of Physics, Loyola University Chicago, Chicago, IL 60660, USA}
\affiliation{Dept. of Physics and Astronomy, University of Canterbury, Private Bag 4800, Christchurch, New Zealand}
\affiliation{Dept. of Physics, University of Maryland, College Park, MD 20742, USA}
\affiliation{Dept. of Astronomy, Ohio State University, Columbus, OH 43210, USA}
\affiliation{Dept. of Physics and Center for Cosmology and Astro-Particle Physics, Ohio State University, Columbus, OH 43210, USA}
\affiliation{Niels Bohr Institute, University of Copenhagen, DK-2100 Copenhagen, Denmark}
\affiliation{Dept. of Physics, TU Dortmund University, D-44221 Dortmund, Germany}
\affiliation{Dept. of Physics and Astronomy, Michigan State University, East Lansing, MI 48824, USA}
\affiliation{Dept. of Physics, University of Alberta, Edmonton, Alberta, T6G 2E1, Canada}
\affiliation{Erlangen Centre for Astroparticle Physics, Friedrich-Alexander-Universit{\"a}t Erlangen-N{\"u}rnberg, D-91058 Erlangen, Germany}
\affiliation{Physik-department, Technische Universit{\"a}t M{\"u}nchen, D-85748 Garching, Germany}
\affiliation{D{\'e}partement de physique nucl{\'e}aire et corpusculaire, Universit{\'e} de Gen{\`e}ve, CH-1211 Gen{\`e}ve, Switzerland}
\affiliation{Dept. of Physics and Astronomy, University of Gent, B-9000 Gent, Belgium}
\affiliation{Dept. of Physics and Astronomy, University of California, Irvine, CA 92697, USA}
\affiliation{Karlsruhe Institute of Technology, Institute for Astroparticle Physics, D-76021 Karlsruhe, Germany}
\affiliation{Karlsruhe Institute of Technology, Institute of Experimental Particle Physics, D-76021 Karlsruhe, Germany}
\affiliation{Dept. of Physics, Engineering Physics, and Astronomy, Queen's University, Kingston, ON K7L 3N6, Canada}
\affiliation{Department of Physics {\&} Astronomy, University of Nevada, Las Vegas, NV 89154, USA}
\affiliation{Nevada Center for Astrophysics, University of Nevada, Las Vegas, NV 89154, USA}
\affiliation{Dept. of Physics and Astronomy, University of Kansas, Lawrence, KS 66045, USA}
\affiliation{Centre for Cosmology, Particle Physics and Phenomenology - CP3, Universit{\'e} catholique de Louvain, Louvain-la-Neuve, Belgium}
\affiliation{Department of Physics, Mercer University, Macon, GA 31207-0001, USA}
\affiliation{Dept. of Astronomy, University of Wisconsin{\textemdash}Madison, Madison, WI 53706, USA}
\affiliation{Dept. of Physics and Wisconsin IceCube Particle Astrophysics Center, University of Wisconsin{\textemdash}Madison, Madison, WI 53706, USA}
\affiliation{Institute of Physics, University of Mainz, Staudinger Weg 7, D-55099 Mainz, Germany}
\affiliation{Department of Physics, Marquette University, Milwaukee, WI 53201, USA}
\affiliation{Institut f{\"u}r Kernphysik, Universit{\"a}t M{\"u}nster, D-48149 M{\"u}nster, Germany}
\affiliation{Bartol Research Institute and Dept. of Physics and Astronomy, University of Delaware, Newark, DE 19716, USA}
\affiliation{Dept. of Physics, Yale University, New Haven, CT 06520, USA}
\affiliation{Columbia Astrophysics and Nevis Laboratories, Columbia University, New York, NY 10027, USA}
\affiliation{Dept. of Physics, University of Oxford, Parks Road, Oxford OX1 3PU, United Kingdom}
\affiliation{Dipartimento di Fisica e Astronomia Galileo Galilei, Universit{\`a} Degli Studi di Padova, I-35122 Padova PD, Italy}
\affiliation{Dept. of Physics, Drexel University, 3141 Chestnut Street, Philadelphia, PA 19104, USA}
\affiliation{Physics Department, South Dakota School of Mines and Technology, Rapid City, SD 57701, USA}
\affiliation{Dept. of Physics, University of Wisconsin, River Falls, WI 54022, USA}
\affiliation{Dept. of Physics and Astronomy, University of Rochester, Rochester, NY 14627, USA}
\affiliation{Department of Physics and Astronomy, University of Utah, Salt Lake City, UT 84112, USA}
\affiliation{Dept. of Physics, Chung-Ang University, Seoul 06974, Republic of Korea}
\affiliation{Oskar Klein Centre and Dept. of Physics, Stockholm University, SE-10691 Stockholm, Sweden}
\affiliation{Dept. of Physics and Astronomy, Stony Brook University, Stony Brook, NY 11794-3800, USA}
\affiliation{Dept. of Physics, Sungkyunkwan University, Suwon 16419, Republic of Korea}
\affiliation{Institute of Physics, Academia Sinica, Taipei, 11529, Taiwan}
\affiliation{Dept. of Physics and Astronomy, University of Alabama, Tuscaloosa, AL 35487, USA}
\affiliation{Dept. of Astronomy and Astrophysics, Pennsylvania State University, University Park, PA 16802, USA}
\affiliation{Dept. of Physics, Pennsylvania State University, University Park, PA 16802, USA}
\affiliation{Dept. of Physics and Astronomy, Uppsala University, Box 516, SE-75120 Uppsala, Sweden}
\affiliation{Dept. of Physics, University of Wuppertal, D-42119 Wuppertal, Germany}
\affiliation{Deutsches Elektronen-Synchrotron DESY, Platanenallee 6, D-15738 Zeuthen, Germany}

\author[0000-0001-6141-4205]{R. Abbasi}
\affiliation{Department of Physics, Loyola University Chicago, Chicago, IL 60660, USA}

\author[0000-0001-8952-588X]{M. Ackermann}
\affiliation{Deutsches Elektronen-Synchrotron DESY, Platanenallee 6, D-15738 Zeuthen, Germany}

\author{J. Adams}
\affiliation{Dept. of Physics and Astronomy, University of Canterbury, Private Bag 4800, Christchurch, New Zealand}

\author[0000-0002-9714-8866]{S. K. Agarwalla}
\altaffiliation{also at Institute of Physics, Sachivalaya Marg, Sainik School Post, Bhubaneswar 751005, India}
\affiliation{Dept. of Physics and Wisconsin IceCube Particle Astrophysics Center, University of Wisconsin{\textemdash}Madison, Madison, WI 53706, USA}

\author[0000-0003-2252-9514]{J. A. Aguilar}
\affiliation{Universit{\'e} Libre de Bruxelles, Science Faculty CP230, B-1050 Brussels, Belgium}

\author[0000-0003-0709-5631]{M. Ahlers}
\affiliation{Niels Bohr Institute, University of Copenhagen, DK-2100 Copenhagen, Denmark}

\author[0000-0002-9534-9189]{J.M. Alameddine}
\affiliation{Dept. of Physics, TU Dortmund University, D-44221 Dortmund, Germany}

\author[0009-0001-2444-4162]{S. Ali}
\affiliation{Dept. of Physics and Astronomy, University of Kansas, Lawrence, KS 66045, USA}

\author{N. M. Amin}
\affiliation{Bartol Research Institute and Dept. of Physics and Astronomy, University of Delaware, Newark, DE 19716, USA}

\author[0000-0001-9394-0007]{K. Andeen}
\affiliation{Department of Physics, Marquette University, Milwaukee, WI 53201, USA}

% ...



\begin{abstract}

IceCube recently reported the observation of TeV neutrinos from the nearby Seyfert galaxy NGC~1068, and the corresponding neutrino flux is significantly higher than the upper limit implied by observations of GeV-TeV gamma rays. This suggests that neutrinos are produced near the supermassive black hole, where the radiation density is high enough to obscure gamma rays.
We use a set of muon neutrinos with interaction vertices inside the detector, which have good sensitivity to sources in the Southern sky, from IceCube data recorded between 2011 and 2021. We then search for individual and collective neutrino signals from 14 Seyfert galaxies in the Southern Sky selected from the Swift Burst Alert Telescope (BAT) AGN Spectroscopic Survey. Using the correlations between keV X-rays and TeV neutrinos predicted by disk-corona models, and assuming production characteristics similar to NGC~1068, a collective neutrino signal search reveals an excess of $6.7_{-3.2}^{+4.0}$ events, which is inconsistent with background expectations at the 3$\sigma$ level of
significance. In this paper, we present new independent evidence that Seyfert galaxies contribute to the extragalactic flux of high-energy neutrinos. 

\end{abstract}


\section{Introduction} \label{sec:intro}

Active galactic nuclei (AGN) have emerged as the leading candidates for the origin of the diffuse high-energy cosmic neutrino flux observed by the IceCube Neutrino Observatory~\citep{Aartsen:2013jdh}. IceCube's observation of this neutrino flux in the TeV to few PeV energy range demonstrates that hadronic interactions contribute considerably to the non-thermal Universe. The high intensity of the neutrino flux at $\sim$30~TeV, observed in two independent detection channels, indicates that high-energy neutrinos originate in dense environments that are opaque to gamma rays~\citep{Senno:2015tra, Murase:2015xka, Bechtol:2015uqb, Capanema:2020rjj, Fang:2022trf}; otherwise, the contribution of the cosmic neutrino sources to the extragalactic gamma-ray background would exceed measurements by the {\em Fermi} Large Area Telescope (LAT)~\citep{Ackermann:2014usa}.  
This picture was reinforced by the observation of neutrino emission from NGC~1068, a nearby Seyfert galaxy~\citep{IceCube:2022der}. NGC~1068 was identified in an IceCube point source search using 8.7 years of muon neutrinos from the Northern Hemisphere. NGC~1068 emerged as the most significant source in the list of sources searched, with a post-trial significance of 4.2$\sigma$. A follow-up analysis with an additional 1.5 years of data found a slightly steeper spectrum ($\gamma=3.3$) and a post-trial significance of 4.3$\sigma$~\citep{Abbasi:2024ofy}. The measured flux of high-energy neutrinos from NGC~1068 was found to be much higher than the gamma-ray flux reported by {\em Fermi}-LAT~\citep{Fermi-LAT:2019yla,Ballet:2020hze} in the GeV regime, and exceeds the upper limits reported by MAGIC~\citep{MAGIC:2019fvw} in the TeV band. The absence of very-high-energy gamma rays from NGC~1068 indicates that the neutrinos are produced in an environment where gamma rays are strongly obscured.

The dense central region of an AGN provides a suitable environment for the efficient production of neutrinos and the suppression of the accompanying gamma-ray signature. Obscuration can also occur at lower photon energies, from the infrared to X-ray bands, due to the dense gas and dust surrounding the AGN~\citep{Padovani:2017zpf}. In the case of NGC~1068, the line of sight passes directly through the dense torus, causing severe X-ray attenuation which qualifies it as a Compton-thick AGN -- a class of objects where even high-energy X-rays are strongly suppressed. 

The AGN corona, a matter-dense and radiation-rich region near the central supermassive black hole, is formed by accretion dynamics and magnetic dissipation~\citep{Miller:1999ix}. The corona is composed of highly magnetized and turbulent plasma~\citep{Inoue:2019fil,Inoue:2019yfs,Murase:2019vdl,Eichmann:2022lxh} and is a candidate for a site of cosmic-ray acceleration.
In the disk-corona model, cosmic-ray acceleration occurs via stochastic acceleration~\citep{Walter:2024pze, Fiorillo:2024akm, Mbarek:2023yeq} and/or magnetic reconnection~\citep{Fiorillo:2023dts}. Cosmic rays interact with the dense gas and intense ambient radiation to produce mesons, among which the charged mesons decay into neutrinos. Motivated by the identification of NGC~1068 as a neutrino source candidate, we focus on the disk-corona model considering stochastic acceleration in the high cosmic ray pressure scenario~\citep{Kheirandish:2021wkm}. NGC~1068 is among the brightest AGN in terms of intrinsic X-ray flux~\citep{Marinucci:2015fqo} and has been modeled to be one of the top candidate Seyfert galaxies that can produce neutrinos detectable by IceCube~\citep{Kheirandish:2021wkm, Murase:2022dog}. We adapt model parameters such that the predicted neutrino flux matches the measured flux from NGC~1068. 

In this work, we search for neutrinos from X-ray bright Seyfert galaxies in the Southern Sky, performing two types of analyses. First, we probe for neutrino emission from each candidate source individually. We consider two scenarios: a generic power-law spectrum and the spectral shape predicted by the disk-corona model. 
Second, we search for cumulative neutrino emission from the selected sources in a stacking search. For the stacking analysis, we test only the disk-corona model. 

\section{Source Candidate Selection} \label{sec:selection}
We select Southern Hemisphere Seyfert galaxies ($\delta<-5\,^{\circ}$) from the 105-month Swift-BAT catalog~\citep{Oh:2018wzc}, avoiding the horizontal region -- which is dominated by atmospheric background contamination -- to optimize neutrino sensitivity for our candidate sources.

\begin{figure}[t]
    \centering
    \includegraphics[width=0.5\linewidth]{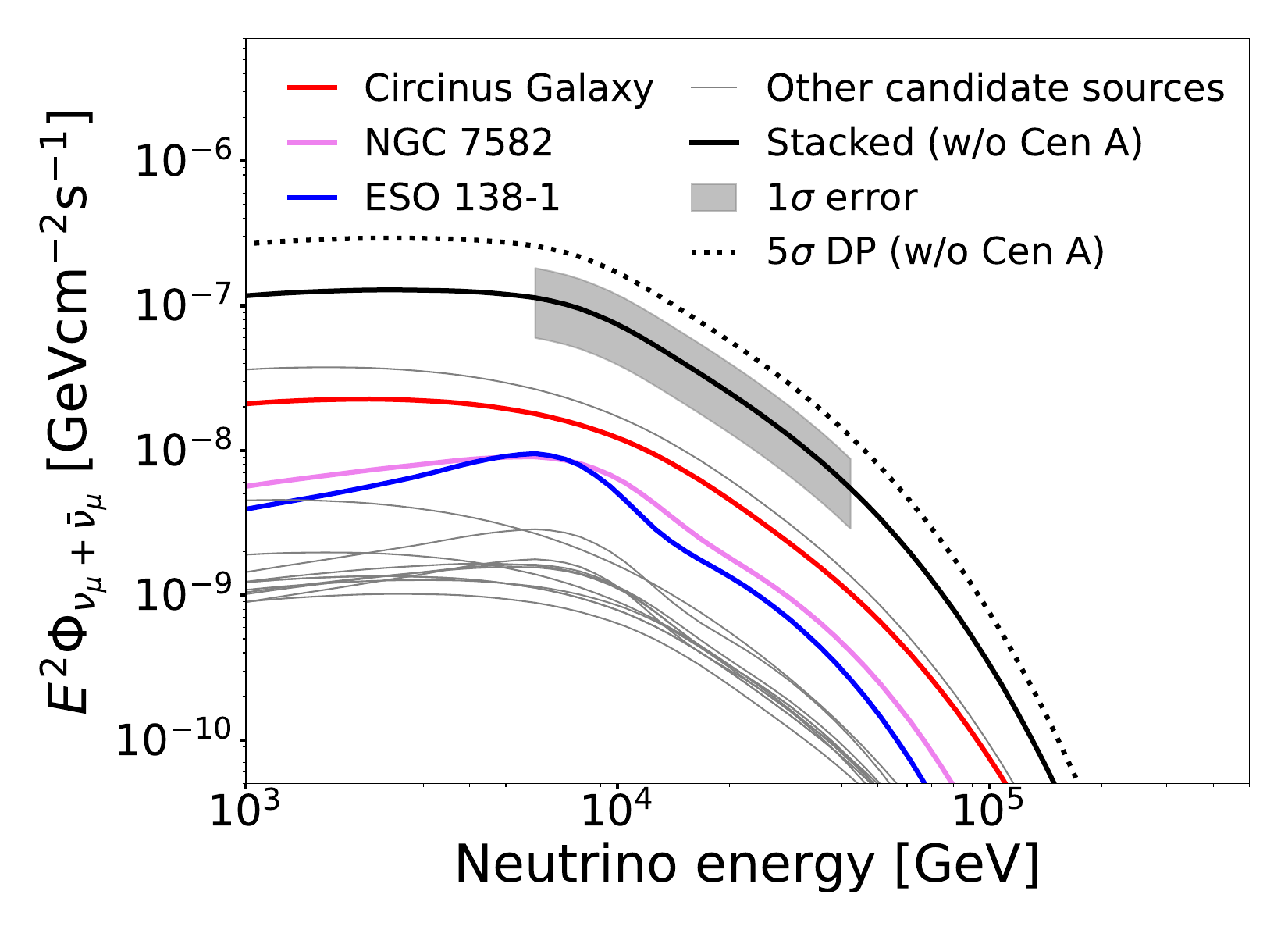}
    \caption{Stacked spectrum according to the best-fit normalization (solid black), with $1\,\sigma$ statistical uncertainty (shaded gray) and the $5\,\sigma$ discovery potential (dotted). The energy range of the gray band is computed from the central 68\% of events contributing to the total TS values. Model predictions of neutrino spectra for the 14 candidate sources~\citep{Kheirandish:2021wkm} are shown as well. The 11 subdominant candidates are shown in gray (see Table~\ref{tab:int_cat}); Centaurus A appears as the highest gray line due to its strong but ambiguous X-ray emission (see Sec.\ref{sec:selection}). The Circinus Galaxy (red), NGC~7582 (light pink), and ESO~138-1 (blue) are the candidates with the largest expected number of signal events, excluding Centaurus A.}\label{fig:stacking_result}
\end{figure}

Given the 2-10 keV absorption-corrected fluxes and distances reported in BASS, we compute the expected neutrino flux from each source using the disk-corona model~\citep{Murase:2019vdl,Kheirandish:2021wkm}. Combined with IceCube’s effective area, we can predict the corresponding number of signal events in the data. For comprehensive coverage, we include candidate sources with expected fluxes slightly below IceCube’s single-source sensitivity threshold, using a threshold of 0.1 predicted signal events per source. These sources and their properties are listed in Table~\ref{tab:int_cat}, ranked by the expected number of signal events, with  Centaurus A and the Circinus Galaxy at the top. An overview of the predicted spectra is given in Figure~\ref{fig:stacking_result}. Centaurus A differs from typical Seyfert galaxies because it has a radio-bright jet, and it is known to be a very-high-energy (VHE) gamma ray emitter~\citep{HESS:2018cxr}. Seyfert galaxies are generally radio-quiet and do not show VHE gamma ray emission~\citep{Fermi-LAT:2011kkd}. Therefore, the origin of the X-ray emission of Centaurus A is still under debate, leading to large uncertainties in the neutrino flux modeling. Hence, this source is analyzed only in isolation, as described in Sec.~\ref{sec:ana_tech}. 

\begin{deluxetable}{lDDDD}[tbh]
\tablecaption{Candidate Source List}
\tablehead{
\colhead{Source} & \multicolumn{2}{c}{Dec [$^\circ$]} & \multicolumn{2}{c}{R.A. [$^\circ$]} & \multicolumn{2}{c}{$\log_{10} (L_{2-10~\mathrm{keV}}^{\rm intr})$} & \multicolumn{2}{c}{$n_{\rm exp}$}
}
\decimals
\startdata
Centaurus A & -43.0 & 201.4 & 42.4 & 2.5 \\
Circinus Galaxy & -65.3 & 213.3 & 42.6 & 1.7 \\
NGC~7582 & -42.4 & 349.6 & 43.5 & 0.8  \\
ESO~138-1 & -59.2 & 252.8 & 44.1 & 0.6 \\
NGC~4945 & -49.5 & 196.4 & 42.1 & 0.2 \\
NGC~424 & -38.1 & 17.9 & 43.8 & 0.2 \\
NGC~4593 & -5.3 & 189.9 & 42.8 & 0.2 \\
MCG-5-23-16 & -30.9 & 146.9 & 43.2 & 0.1 \\
NGC~3783 & -37.7 & 174.8 & 43.4 & 0.1 \\
IC4329A & -30.3 & 207.3 & 43.8 & 0.1  \\
NGC~4507 & -39.9 & 188.9 & 43.5 & 0.1 \\
NGC~5728 & -17.3 & 220.6 & 42.9 & 0.1 \\
NGC~5643 & -44.2 & 218.2 & 42.4 & 0.1 \\
NGC~3081 & -22.8 & 149.9 & 42.7 & 0.1 \\
\enddata
\tablecomments{
Details of the candidate source list. Logarithm of intrinsic deabsorbed 2-10~keV X-ray luminosity, $\log_{10}(L_{2-10~\mathrm{keV}}^{\rm intr} /\mathrm{erg\,cm^{-2}\,s^{-1}})$ are from BASS. Numbers of expected signal events ($n_{\rm exp}$) are predicted by the disk-corona model.
}
\label{tab:int_cat}
\end{deluxetable}

\section{IceCube detector and data sample}\label{sec:detector}

The IceCube Neutrino Observatory, located deep under the ice at the South Pole, is a cubic-kilometer Cherenkov neutrino telescope. 
The detector is composed of an array of 5,160 digital optical modules (DOMs), each equipped with a photomultiplier tube (PMT) and on-board read-out electronics~\citep{IceCube:2016zyt}. 
Interactions of neutrinos with nucleons and electrons in the ice or underlying bedrock produce relativistic charged particles, which emit Cherenkov photons collected by the PMTs. The photon arrival times recorded by each DOM are used to reconstruct the energy and direction of each event. The flavor of neutrino and type of neutrino interaction determine the event morphology in the detector. Muon tracks from $\nu_{\mu}/\bar{\nu}_{\mu}$ charged-current interactions provide the best angular resolution and thus are commonly chosen in our searches for point-like neutrino emission~\citep{IceCube:2016zyt}.

However, cosmic-ray interactions in Earth's atmosphere also produce large numbers of muons and neutrinos. Because of their track-like signature, they pose a background in searches for astrophysical neutrinos. Atmospheric muons trigger the detector at a rate of 3~kHz~\citep{IceCube:2015wro}. They mostly impact the Southern Sky, since the muons produced in the Northern Sky are effectively blocked by the Earth. For the analysis presented here, we rely on an event selection technique that suppresses the atmospheric backgrounds for down-going track events originating from the Southern Hemisphere. The method relies on selecting tracks that start in the detector (``starting tracks"), i.e., the neutrino interacts inside the detector. A dedicated vetoing technique separates neutrino candidate events from atmospheric muons. By efficiently rejecting atmospheric muons, the selection also benefits from a self-veto effect~\citep{Gaisser:2014bja} which results in a suppression of atmospheric neutrinos, since they are accompanied by muons from the same cosmic-ray air shower. 
The corresponding dataset, Enhanced Starting Track Event Selection (ESTES), contains events collected by IceCube over $\sim$10 years from May 2011 to January 2022. This enhanced selection employs a dynamic veto and machine learning to significantly improve the retention of starting track events in the Southern Sky, achieving a median angular resolution of 1.4$^\circ$ and targeting neutrino energies of 1–500 TeV. More details about the reconstruction and selection techniques used to create the ESTES sample can be found in~\citet{IceCube:2024fxo,IceCube:2025zyb}.
In total $10,350$ events pass all selection criteria, of which $2,091$ are down-going and therefore contribute to this analysis. 

\begin{figure}[t]
    \centering
    \includegraphics[width=0.45\linewidth]{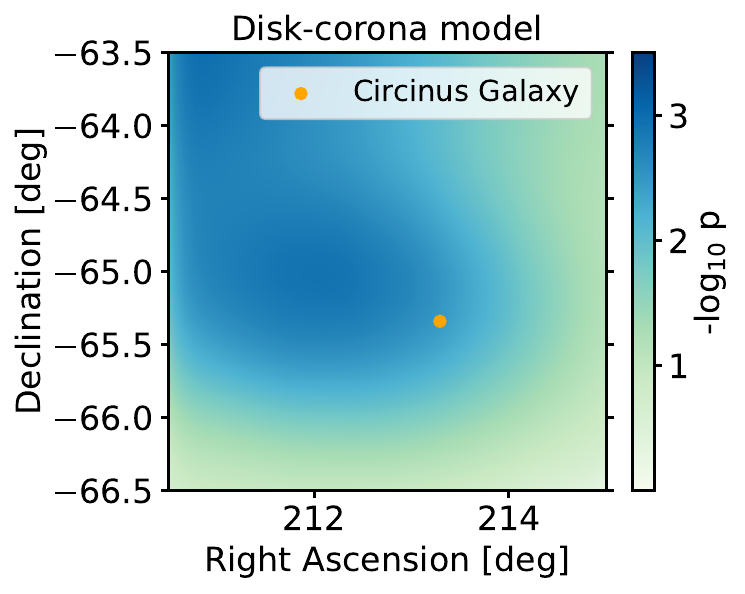}
    \includegraphics[width=0.45\linewidth]{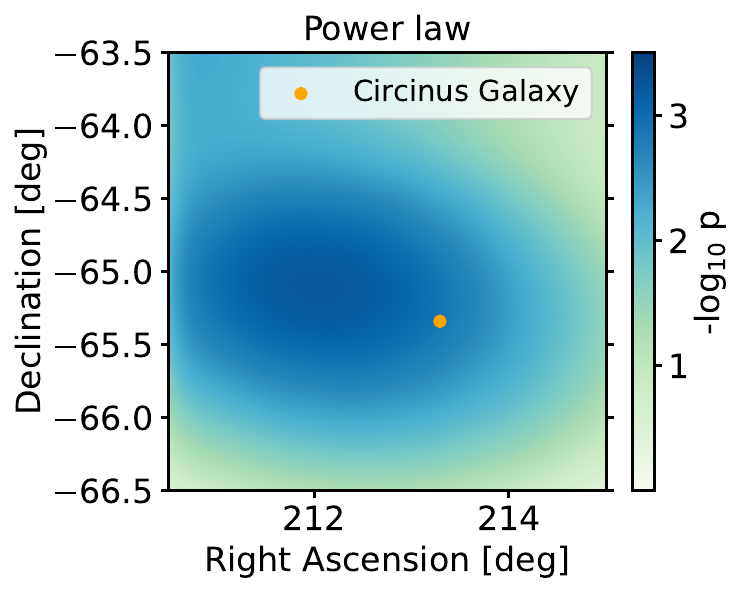}
    \caption{Local (pre-trial) $p$-value maps near the most significant source - the Circinus Galaxy (orange), assuming the disk-corona model spectrum (left) and the power-law spectrum (right).}
    \label{fig:p_scan}
\end{figure}

\section{Analysis techniques}\label{sec:ana_tech} 
We perform two analysis approaches on the selected candidate sources (see Sec.~\ref{sec:selection}): an individual candidate source search and a stacking search. In the individual source search, we search for an excess of neutrinos from the direction of each candidate source.
This is done separately for two model tests: the disk-corona model and the generic power-law spectrum. 
In the stacking search, we study the accumulated emission from the selected sources. Each source is weighted by the expected number of signal events in the detector, where the expectation is computed using the disk-corona model and the distance of the source. We do not include Centaurus A in the stacking search because of the large uncertainties in the model prediction due to the ambiguous origin of its X-ray emission (see Sec.~\ref{sec:selection}). Centaurus A would otherwise receive the strongest weight and dominate this stacking analysis (see Figure~\ref{fig:stacking_result}).
To test a possible astrophysical neutrino signal against the background-only hypothesis, we rely on an unbinned likelihood ratio test ~\citep{Braun:2008bg}. The test incorporates the energy, direction, and the directional uncertainty of each event in the data sample. 

Our modeling of the predominant background components (atmospheric muons, atmospheric neutrinos, and the isotropic contribution from unresolved sources) relies on the experimental data, which is dominated by these backgrounds. Following established IceCube methods~\citep{IceCube:2025zyb}, we construct the background probability density function (PDF) by randomizing the right ascension of the events. This provides a good approximation for these background events since, due to IceCube's location, they are uniform in right ascension.
However, IceCube recently reported evidence of neutrino emission from the Galactic plane \citep{IceCube:2023ame}.
This adds a new background component to our search.
Without modifications, the standard randomizing method would over-estimate the background for Seyfert galaxies far from the Galactic plane and under-estimate the background for Seyfert galaxies close to the Galactic plane due to this background contamination. For a similar reason, to prevent potential signal events from the brightest candidate sources from being randomized into the background, we remove events near their locations. Hence, in the randomization process, we modified the standard scrambling method as follows. 
Events in the vicinity of the brightest (according to model expectation) Seyfert galaxies (Centaurus A, Circinus, NGC~7582, and ESO~138-1, radius $7^{\circ}$) or the Galactic plane (absolute value of Galactic latitude $<10^{\circ}$ ) are removed from the dataset. 
The remaining events are then randomized in right ascension. Since some background events are removed in the previous step, we randomly resample from the remaining events to compensate for those missing due to masking, thereby preserving the total number of background events. 
This oversampling step approximates the atmospheric flux and the isotropic astrophysical background, uniform in right ascension, as before. 
However, by construction, it does not account for the contribution from the Galactic plane. To model the Galactic plane, we use Monte Carlo (MC)–generated events.
Here we assume IceCube's best estimate of the Galactic neutrino flux, using {\em Fermi}-LAT gamma-ray measurements based on the $\pi^0$ template~\citep{Fermi-LAT:2012edv} for the spatial distribution of neutrinos, and a power-law with a spectral index of 2.7 and normalization of 21.8$\times10^{–12}$ TeV cm$^{–2}$ s$^{–1}$ at 100 TeV for the flavor-averaged flux of neutrinos multiplied by energy squared~\citep{IceCube:2023ame}. 
This technique results in hybrid data + MC background-only datasets that represent the known properties of all relevant background components.
Using these datasets, we compute the background PDFs used in the likelihood calculation. 
The hybrid datasets are also used to derive the distributions of the maximum likelihood test statistic under the background-only hypothesis, a crucial ingredient for the likelihood ratio test.

We tested the performance of the analysis by simulating neutrino emission from the selected Seyfert galaxies in addition to the backgrounds. 
We find that the stacking analysis, using neutrino expectations calculated from the disk-corona model, would yield a $\geq5\sigma$ discovery in $50\%$ of the simulations (discovery potential) if a total of 15.3 signal neutrinos are distributed across the sources.
The disk-corona model with our choice of parameters predicts $4.7$ events, corresponding to an expected significance of $2.1\sigma$. Figure~\ref{fig:stacking_result} compares the modeled energy spectrum to the 5$\sigma$ discovery potential based on the disk-corona model.  

\begin{figure}[t!]
    \centering
    \includegraphics[width=0.5\linewidth]{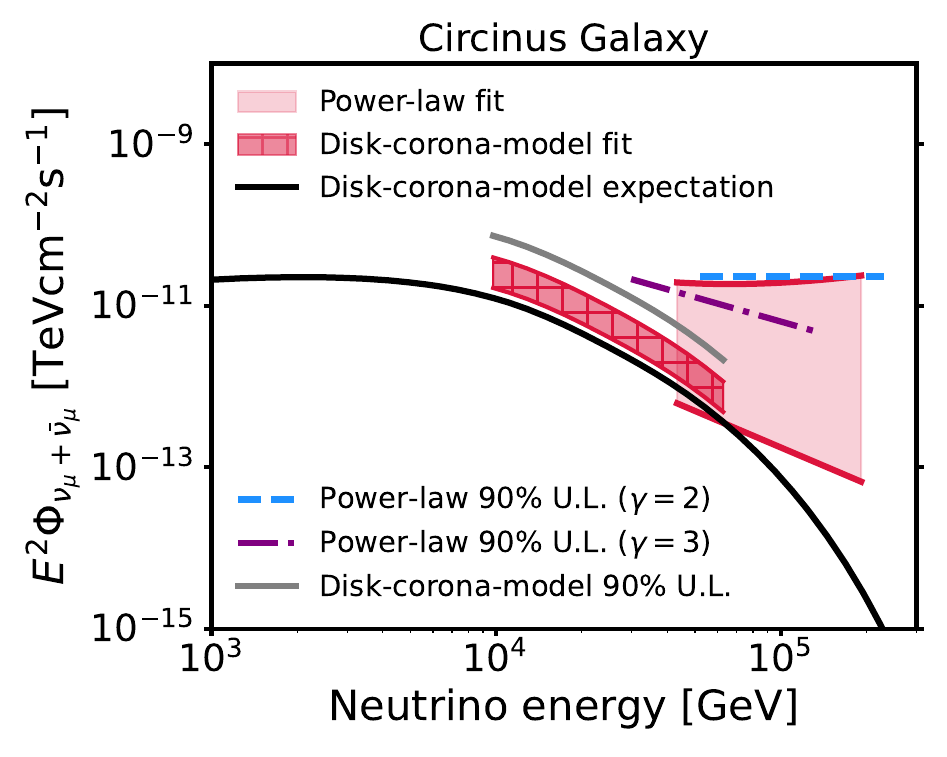}
    \caption{The predicted neutrino spectrum for the Circinus Galaxy from the disk–corona model (solid black) is compared to the 68\% uncertainty regions of the best-fit spectra assuming the disk–corona scenario (hatched) and a power law (shaded). The energy range corresponds to the central 68\% of events contributing to the total TS values. The 90\% confidence level upper limits of power-law analysis for $E^{-2}$ (dashed blue) and $E^{-3}$ (dash-dotted purple) fluxes, as well as the model analysis (solid gray) are shown. The post-trial statistical significance for the Circinus Galaxy is 1.8$\sigma$. }
    \label{fig:circinus_flux}
\end{figure}

\section{Results} \label{sec:result} 

\begin{deluxetable*}{lccccccc}[hbt!]
\tablecaption{Summary of Brightest Individual Sources and Stacking Analysis Results}
\tablewidth{28pt}
\tablehead{
\colhead{} & \colhead{$n_{\rm exp}$} & \colhead{TS} & \colhead{$\hat{n}_{\rm s}$} & \colhead{$\hat{\gamma}$} & \colhead{$p_{\rm pre}$} & \colhead{$p_{\rm post}$} & \colhead{90\% U.L.}}
\decimals
\startdata
Disk-corona & \phantom{$-$} & \phantom{$-$} & \phantom{$-$} & \phantom{$-$} & \phantom{$-$} & \phantom{$-$} & \hspace{0.1cm}$n_{\rm{s}}$ \\ \hline
Circinus Galaxy & 1.7  & 6.7 & 3.6 & $-$ & $0.003\,(2.7\,\sigma)$ & $-$ & 10.0 \\
ESO~138-1 & 0.6  & 3.0  & 1.7  & $-$ & $0.03\,(1.9\,\sigma)$ & $-$ & 5.7 \\
NGC~7582  & 0.8 & 1.4 & 1.1 & $-$ & $0.05\,(1.6\,\sigma)$ & $-$ & 5.1 \\ \hline
Power-law & \phantom{$-$} & \phantom{$-$} & \phantom{$-$} & \phantom{$-$} & \phantom{$-$} & \phantom{$-$} & $\phi(\rm{TeV}^{-1} \rm{cm}^{-2} \rm{s}^{-1} \text{at 1\ TeV}$) \\ \hline
Circinus Galaxy  & $-$ & 10.4 & 3.1 & 2.5 & $0.001\,(3.1\,\sigma)$ & $0.033\,(1.8\,\sigma)$ & $6.38 \times 10^{-10}$ \\
NGC~7582 & $-$ & 1.7 & 1.7 & 4.0 & $0.05\,(1.6\,\sigma)$ & $-$ & $2.56 \times 10^{-10}$ \\
ESO~138-1  & $-$ & 3.0  & 1.9 & 3.6 & $0.06\,(1.6\,\sigma)$ & $-$ & $2.97 \times 10^{-10}$ \\ \hline
Stacked sources & \phantom{$-$} & \phantom{$-$} & \phantom{$-$} & \phantom{$-$} & \phantom{$-$} & \phantom{$-$} & \hspace{0.1cm}$n_{\rm{s}}$ \\ \hline
13 sources w/o CenA  & 4.7 & 7.8 & 6.7 & $-$ & $0.0013\,(3.0\,\sigma)$ & $-$ & 14.3 \\
\enddata
\tablecomments{
Summary of stacking search results with disk-corona model assumptions and individual source search results for the three most significant sources: Circinus Galaxy, NGC~7582, and ES~138-1, with both disk-corona model and power-law spectral assumptions. The best-fit test-statistic (TS), number of signal events ($\hat{n}_{\rm s}$), pre-trial $p$-value ($p_{\rm pre}$), post-trial $p$-value ($p_{\rm post}$), and 90\% confidence level upper limit are listed for both analyses ($E^{-3}$ flux for power-law analysis and number of signal neutrinos for the model analysis). For the model analysis, the number of expected signal neutrinos ($n_{\rm exp}$) is listed while for the power-law analysis the best-fit spectral index ($\hat{\gamma}$) is shown.
}
\label{tab:results}
\end{deluxetable*}

Assuming the disk-corona model spectrum, the stacking analysis results in a combined best-fit excess of $6.7_{-3.2}^{+4.0}$ signal events over background and yields a significance of 3.0$\sigma$ (single-sided $p$-value = $1.3\times 10^{-3}$); see Table~\ref{tab:results} for details. This is consistent with the model prediction of 4.7 events. We determine the 68\% central confidence interval for the number of signal events $n_s$  from the likelihood function assuming Wilks' theorem~\citep{Wilks:1938dza}. The shaded gray band in Figure~\ref{fig:stacking_result} shows the measured flux for the stacking search with uncertainties, compared with the 5$\sigma$ discovery potential.
We also show the predicted spectra of the individual candidate sources.

We also studied each candidate source individually, testing both spectral hypotheses. We find a non-zero, albeit not statistically significant, number of signal neutrinos from the locations of the Circinus Galaxy, ESO~138-1, and NGC~7582. The most significant result is obtained in the direction of the Circinus Galaxy, corresponding to a local (pre-trial) significance of 3.1$\sigma$ with a best-fit spectral index of 2.5, assuming the neutrino spectrum follows a power-law function. After accounting for the number of candidate sources in the list which is 14 and the two spectral assumptions, the post-trial significance of the single brightest candidate source is $1.8\sigma$ (p-value = 0.033). The number of 3.1 best-fit signal neutrinos identified in our analysis is not statistically significant, so we place an upper limit on the neutrino flux from both the Circinus Galaxy and the other sources in this study. For each limit we assume a power-law with spectral index of $\gamma=3$. A summary of these results can be found in Table~\ref{tab:results}. We performed a spatial scan of the local $p$-value around the direction of the Circinus Galaxy. The result is shown in Figure~\ref{fig:p_scan}. 
The best-fit and 90\% confidence level upper limit fluxes of both disk-corona and power-law assumptions are shown in Figure~\ref{fig:circinus_flux} along with the expected spectrum of the model. 
\begin{figure}[bth]
    \centering
    \includegraphics[width=0.6\linewidth]{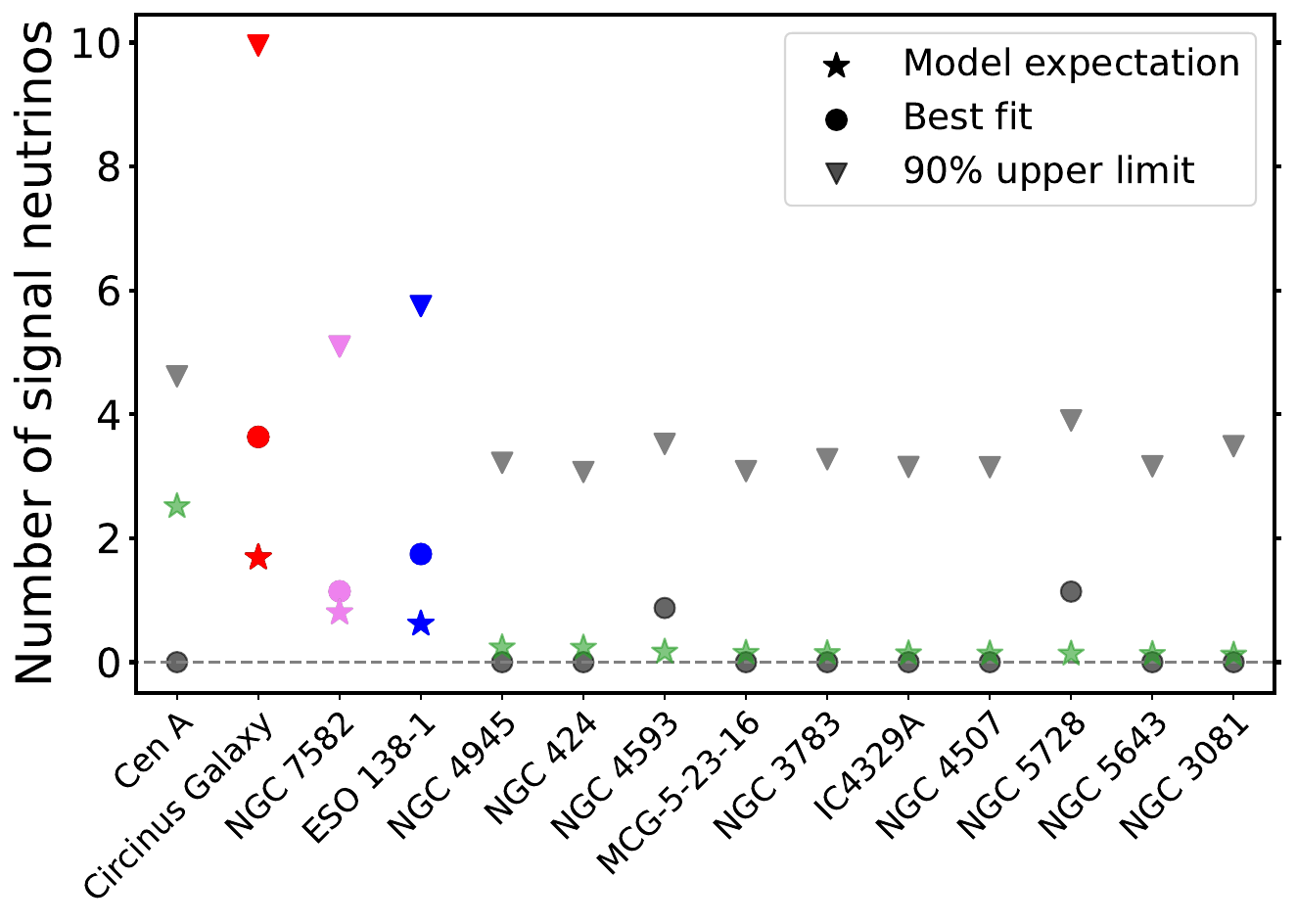}
    \caption{Expected (star), best-fit (dot), and 90\% confidence level upper limit (triangle) number of signal neutrinos assuming disk-corona model spectra, with the three most significant sources -- Circinus galaxy (red), NGC~7582 (pink), and ESO~138-1 (blue) -- highlighted. Sources from left to right have a decreasing number of expected signal neutrinos. The ambiguous origin of the X-ray emission from Centaurus A introduces large uncertainties in the expected neutrino flux; therefore, we include it in the individual source search but not in the weighted stacking analysis.}
    \label{fig:catalog_result}
\end{figure}
Assuming the disk-corona model, Figure~\ref{fig:catalog_result} summarizes the expected number of neutrinos, the best-fit number of signal neutrinos ($n_s$), and the upper limits (90\% C.L.) for all the candidate sources. It demonstrates that the 3.0$\sigma$ result obtained in the stacking search is due to the combined excess of the Circinus Galaxy, ESO~138-1, and NGC~7582, for which the best-fit neutrino counts are correlated with predictions from the disk-corona model. Finally, more details about the individual results for all 14 candidate sources in our list, including flux upper-limits, can be found in Table~\ref{tab:more_results} in the Appendix. 

Before unblinding the experimental data, we performed extensive validation of all methods, in particular the improved background estimation and background sampling techniques. We verified post-unblinding that the traditional method, which leads to a small over-estimation of the background in our stacking search, would have yielded a similar result. Specifically, had we not used the source masking and Galactic plane modeling techniques, we would have obtained a slightly reduced significance of $2.8\sigma$, consistent with the corresponding reduction in the analysis' sensitivity.

\section{Summary and Discussion}\label{sec:summary} 
\begin{deluxetable*}{lcccc}[bth!]
\tablecaption{Summary of Searches of X-ray Bright Seyfert Galaxies and Hard X-ray Bright AGNs using IceCube Data}
\tablewidth{10pt}
\tablehead{
\multicolumn{1}{c}{source}  & \multicolumn{1}{c}{$\sigma_{\rm post}$} & \multicolumn{1}{c}{$n_{s}$}  &\multicolumn{1}{c}{flux [$\times 10^{-11}$ TeV$^{-1}$cm$^{-2}$s$^{-1}$]}  & \multicolumn{1}{c}{data sample}
}
\decimals
\startdata
~\citep{IceCube:2022der}& & & & \\ NGC~1068 & $4.2\,\sigma$ & $-$ &
$5.0^{+1.5}_{-1.5}$ (best-fit) 
& May 2011 to May 2020 (NT)\\ \hline
~\citep{PhysRevLett.132.101002}& &  &  &\\ NGC~3079 & $\sim3\,\sigma$ & $-$ & $3.2^{+4.0}_{-2.5}$ (best-fit) & Apr 2008 to Jul 2018 (PS)\\ \hline
~\citep{Abbasi:2024hwv} & & & &  \\ NGC~4151 & $2.9\,\sigma$ & $-$ & $1.51^{+0.99}_{-0.81}$ (UL) & Apr 2008 to May 2020 (PS)\\ \hline
\citep{Abbasi:2024ofy} &  & & & \\
CGCG~420-015~ & $2.3\,\sigma$ & 46.4 (UL) & $-$ &May 2011 to Feb 2022 (NT)\\
binomial test & $2.7\,\sigma$ & $-$ &$-$\\\hline
\citep{2025evidenceseyfert} & &  &  \\
binomial test & $3.3\,\sigma$ & $-$ & $-$ &May 2011 to Nov 2023 (NT) \\\hline
(\textit{this work}) & & & & \\Circinus Galaxy & $1.8\,\sigma$ & $-$ & 63.8 (UL)  & May 2011 to Jan 2022 (ST)\\
stacking w/o CenA & $3.0\,\sigma$ & 14.3 (best-fit) &$-$\\
\enddata
\tablecomments{
A summary of the datasets and the key results from the searches for hard X-ray bright AGNs \citep{Abbasi:2024hwv}, X-ray bright Seyfert galaxies \citep{PhysRevLett.132.101002,Abbasi:2024ofy,2025evidenceseyfert} using the all-sky point source data (PS) ~\citep{2020PhRvL.124e1103A} or the Northern Sky through-going tracks (NT)~\citep{IceCube:2022der}, and this work, using the Southern Sky starting tracks (ST)~\citep{IceCube:2024fxo}. The post-trial significance ($\sigma_{\rm post}$) for each search is listed. The 90\% confidence level upper limits (UL) are reported for an $E^{-3}$ spectrum for the power-law results, and as the number of signal (ns) neutrinos for the results assuming disk-corona model. The most significant individual sources identified in each search are NGC~3079, NGC~4151, CGCG~420-015, and the Circinus Galaxy. The results of the collective searches for neutrino emission--assuming the disk-corona model spectrum for each candidate source--are also included, both with (stacking analysis) and without (binomial test) assumptions about the relative contributions of the sources.
}
\label{tab:three_results}
\end{deluxetable*}

In the Southern Sky, even though our studies do not identify significant neutrino emission from individual sources, the stacking result is inconsistent with the background hypothesis at $3.0\sigma$ significance, driven by five galaxies--led by three nearby AGNs in the Circinus Galaxy, NGC~7582, and ESO~138-1. Within statistical fluctuations, our findings are consistent with the predictions of the disk-corona model.
This work provides evidence independent of NGC~1068 that some X-ray bright Seyfert galaxies are sources of high-energy neutrinos. Table~\ref{tab:three_results} summarizes the searches in both hemispheres.

In this work, we studied 14 candidate sources, selected from the BASS AGN survey, for which the models of neutrino production in the AGN coronae predict the strongest neutrino signal. Our study benefits from the new event selection techniques (ESTES) optimized for isolating starting track events \citep{IceCube:2024fxo,IceCube:2025zyb}, thereby efficiently suppressing the large background from atmospheric muons, which traditionally limit searches for neutrino sources below 100~TeV in the Southern sky.
In addition to the generic power-law spectrum, we analyzed each candidate source assuming the spectrum follows the prediction of the disk-corona model. 
We also searched for the aggregated neutrino signal using a stacking analysis (excluding Centaurus A) based on the same model. 
The latter analysis identified a total excess of $6.7_{-3.2}^{+4.0}$ events, consistent with the model prediction of $4.7$ events. 
This result is inconsistent with the background hypothesis at $3.0\sigma$ significance. 

In the Northern sky, motivated by the identification of neutrino emission from the Seyfert galaxy NGC~1068 \citep{IceCube:2022der}, IceCube has searched for neutrinos from X-ray bright Seyfert galaxies~\citep{Abbasi:2024ofy} using the same event selection but improved calibration and reconstruction and an additional 1.7 years of data.
28 sources were selected from the BASS catalog requiring their 2~keV - 10~keV intrinsic X-ray flux to be at least 10\% of that of NGC~1068. Although the stacking search in the Northern Sky did not find statistically significant collective emission from the candidate sources, a binomial test--assuming disk-corona model spectra for the stacked sources (excluding NGC~1068) and no assumption about their relative contributions--yielded a post-trial significance of $2.7\sigma$.
Another analysis, using the same dataset described in \cite{IceCube:2022der} along with two additional years of data, was conducted around the same time as \cite{Abbasi:2024ofy}. 
Similar to \cite{Abbasi:2024ofy}, \cite{Abbasi:2024hwv} reported no significant collective emission from the stacking search. In the individual source search, \cite{Abbasi:2024hwv} selected the top-ranked AGNs from the BASS catalog by comparing the ratio of their intrinsic 14–195 keV X-ray flux to the neutrino sensitivity, retaining sources with a ratio no more than 10 times smaller than that of the top-ranked source. In total, 43 candidates were selected. Most of the candidates are located in the Northern sky due to better sensitivity with the chosen through-going track dataset. In this search, NGC~4151 was the most significant source, with a post-trial significance of 2.9$\sigma$. More recently, IceCube reported observing a 3.3$\sigma$ excess from 11 sources out of 47 selected X-ray bright Seyfert galaxies that were not included in the list
of gamma-ray emitters (excluding NGC~1068) using 13.1 years of through-going track events. 
In addition to these IceCube Collaboration results, we note that independent researchers analyzing 10 years of publicly available IceCube data~\citep{2020PhRvL.124e1103A} have reported a $\sim$3~$\sigma$ excess associated with NGC~3079~\citep{PhysRevLett.132.101002}.

The current analysis can be improved in the future by including a sample of cascade-type events (mostly from high-energy electron neutrino interactions) detected by IceCube~\citep{IceCube:2023ame} that provide competitive sensitivity to these interesting sources. However, a firm discovery of this potential class of neutrino emitters will benefit from future neutrino detectors with larger effective areas and improved angular resolution, such as Baikal-GVD, KM3NeT, P-ONE, NEON, TRIDENT, HUNT, and IceCube-Gen2~\citep{Avrorin:2013uyc,Adrian-Martinez2016,P-ONE:2020ljt,Zhang_2025,TRIDENT:2022hql,Huang:2023mzt,IceCube-Gen2:2020qha}. 

\section*{Acknowledgements} 
The IceCube collaboration acknowledges the significant contributions to this manuscript from Ali Kheirandish, Qinrui Liu, and Shiqi Yu. 
The authors gratefully acknowledge the support from the following agencies and institutions:
USA {\textendash} U.S. National Science Foundation-Office of Polar Programs,
U.S. National Science Foundation-Physics Division,
U.S. National Science Foundation-EPSCoR,
U.S. National Science Foundation-Office of Advanced Cyberinfrastructure,
Wisconsin Alumni Research Foundation,
Center for High Throughput Computing (CHTC) at the University of Wisconsin{\textendash}Madison,
Open Science Grid (OSG),
Partnership to Advance Throughput Computing (PATh),
Advanced Cyberinfrastructure Coordination Ecosystem: Services {\&} Support (ACCESS),
Frontera and Ranch computing project at the Texas Advanced Computing Center,
U.S. Department of Energy-National Energy Research Scientific Computing Center,
Particle astrophysics research computing center at the University of Maryland,
Institute for Cyber-Enabled Research at Michigan State University,
Astroparticle physics computational facility at Marquette University,
NVIDIA Corporation,
and Google Cloud Platform;
Belgium {\textendash} Funds for Scientific Research (FRS-FNRS and FWO),
FWO Odysseus and Big Science programmes,
and Belgian Federal Science Policy Office (Belspo);
Germany {\textendash} Bundesministerium f{\"u}r Forschung, Technologie und Raumfahrt (BMFTR),
Deutsche Forschungsgemeinschaft (DFG),
Helmholtz Alliance for Astroparticle Physics (HAP),
Initiative and Networking Fund of the Helmholtz Association,
Deutsches Elektronen Synchrotron (DESY),
and High Performance Computing cluster of the RWTH Aachen;
Sweden {\textendash} Swedish Research Council,
Swedish Polar Research Secretariat,
Swedish National Infrastructure for Computing (SNIC),
and Knut and Alice Wallenberg Foundation;
European Union {\textendash} EGI Advanced Computing for research;
Australia {\textendash} Australian Research Council;
Canada {\textendash} Natural Sciences and Engineering Research Council of Canada,
Calcul Qu{\'e}bec, Compute Ontario, Canada Foundation for Innovation, WestGrid, and Digital Research Alliance of Canada;
Denmark {\textendash} Villum Fonden, Carlsberg Foundation, and European Commission;
New Zealand {\textendash} Marsden Fund;
Japan {\textendash} Japan Society for Promotion of Science (JSPS)
and Institute for Global Prominent Research (IGPR) of Chiba University;
Korea {\textendash} National Research Foundation of Korea (NRF);
Switzerland {\textendash} Swiss National Science Foundation (SNSF).\\

\appendix

\begin{deluxetable}{lDDDD|DDDDD}[tbh]
\tablecaption{Results of individual candidate source searches.}
\tablewidth{26pt}
\tablehead{
\colhead{} & \multicolumn{8}{c}{Disk-Corona Model} & \multicolumn{8}{c}{Power-Law} \\
\colhead{Source} & \multicolumn{2}{c}{$n_{\rm exp}$} & \multicolumn{2}{c}{$\hat{n}_{\rm s}$} & \multicolumn{2}{c}{$-\log_{10} p$} & \multicolumn{2}{c}{$n_{\rm UL}$} & \multicolumn{2}{c}{$\hat{n}_s$} & \multicolumn{2}{c}{$\hat{\gamma}$} & \multicolumn{2}{c}{$-\log_{10}p$} & \multicolumn{2}{c}{$\phi^{E^{-2}}_{90\%}$} & \multicolumn{2}{c}{$\phi^{E^{-3}}_{90\%}$}
}
\decimals
\startdata
Centaurus A & 2.5 & 0.0 & 0.0 & 4.6 & 0.0 & 4.0 & 0.0 & 1.0 & 22.1 \\
Circinus Galaxy & 1.7 & 3.6 & 2.5 & 10.0 & 3.1 & 2.5 & 2.9 & 2.3 & 63.8 \\
NGC~7582 & 0.8 & 1.1 & 1.3 & 5.1 & 1.7 & 4.0 & 1.3 & 1.1 & 25.6 \\
ESO~138-1 &  0.6 & 1.7 & 1.5 & 5.7 & 1.9 & 3.6 & 1.2 & 1.1 & 29.7 \\
NGC~4945 & 0.2 & 0.0 & 0.0 & 3.2 & 0.0 & 4.0 & 0.0 & 0.6 & 15.8 \\
NGC~424 &  0.2 & 0.0 & 0.0 & 3.1 & 0.0 & 2.2 & 0.0 & 0.6 & 14.8 \\
NGC~4593 &0.2 & 0.9 & 0.6 & 3.5 & 0.6 & 2.9 & 0.4 & 0.6 & 8.9 \\
MCG-5-23-16 & 0.1 & 0.0 & 0.0 & 3.1 & 1.0 & 1.7 & 1.0 & 1.0 & 22.6 \\
NGC~3783 & 0.1 & 0.0 & 0.0 & 3.3 & 0.0 & 4.0 & 0.0 & 0.7 & 15.8 \\
IC4329A & 0.1 & 0.0 & 0.0 & 3.1 & 0.0 & 3.8 & 0.0 & 0.7 & 15.3 \\
NGC~4507 & 0.1 & 0.0 & 0.0 & 3.1 & 0.0 & 3.8 & 0.0 & 0.7 & 15.1 \\
NGC~5728 & 0.1 & 1.1 & 0.9 & 3.9 & 1.9 & 4.0 & 0.9 & 0.9 & 16.9 \\
NGC~5643 & 0.1 & 0.0 & 0.0 & 3.2 & 0.0 & 4.0 & 0.0 & 0.6 & 15.2 \\
NGC~3081 & 0.1 & 0.0 & 0.0 & 3.5 & 0.0 & 4.0 & 0.0 & 0.8 & 16.1 \\
\enddata
\tablecomments{
Details of candidate source list and results. Values of best-fit $\hat{n}_s$ (and $\hat{\gamma}$) and pre-trial $p$-values using disk-corona model (power-law) spectral assumption are shown. Numbers of expected signal events ($n_{\rm exp}$) and 90\% upper limit ($n_{\rm UL}$) are listed for results using model assumptions. For the power-law analysis, 90\% upper limit fluxes are listed, parameterized as $\phi^{E^{-\gamma}}_{90\%} \left(E/1\,\mathrm{TeV}\right)^{-\gamma} \times 10^{-11}\,\mathrm{TeV^{-1}\,cm^{-2}\,s^{-1}}$, where $\gamma=2$ or 3.
}
\label{tab:more_results}
\end{deluxetable}

\bibliography{main}

\end{document}